# Pressure dependent structure of neat liquid methanol, CH$_3$OH: molecular dynamics simulations with various united atom type potentials

Imre Bakó[1], László Pusztai[2,3], Orest Pizio[4]

[1] Research Centre for Natural Sciences, H-1117 Budapest, Magyar tudósok körútja 2., Hungary

[2] Wigner Research Centre for Physics, H-1121 Budapest, Konkoly Thege M. út 29-33., Hungary

[3] Faculty for Advanced Science and Technology (FAST), Kumamoto University, 2-39-1 Kurokami, Chuo-ku, Kumamoto, 860-8555, Japan

[4] Instituto de Química, Universidad Nacional Autónoma de México, Circuito Exterior, 04510, Ciudad de México, México

**Abstract**

Molecular dynamics computer simulations have been conducted on neat liquid methanol, using three different 'united atom' (three site) interatomic potentials: TraPPE [*J. Phys. Chem. B* 105, 3093 (2001)], UAM-I [*J. Mol. Liq.* 323, 114576 (2021)] and OPLS/2016 [*J. Chem. Phys.* 145, 034508 (2016)]. The effects of pressure, between 1 bar and 6 kbar, have been evaluated on total scattering structure factors, partial radial distribution functions, as well as on collective characteristics such as ring size distributions and cluster size distributions. Agreement with experimental density is nearly quantitative for all the three force fields, and major trends observed for recent pressure dependent neutron diffraction data are reproduced qualitatively. The OPLS/2016 force field, in general, generates properties that are markedly different from results originating from the other potentials. Pressure effects are hardly noticeable on most partial radial distribution functions and on the distribution of the number of hydrogen bonded neighbours. On the other hand, collective structural properties, like cluster- and ring size distributions exhibit significant changes with increasing pressure: larger clusters become more numerous, whereas the number of cyclic clusters, i.e., rings, decrease. The self-diffusion coefficient decreases with increasing pressure, and the same is valid for the average lifetime of hydrogen bonds.

**Keywords**: liquid methanol, structure, pressure dependence, molecular dynamics

Corresponding author: PUSZTAI, László (pusztai.laszlo@wigner.hun-ren.hu)

# Introduction

It is our great pleasure to be able to contribute to the Carlos Vega Festschrift. Professor Vega has devoted much of his scientific career to computer simulation investigations concerning water, including development of the highly successful interatomic potential TIP4P/2005 [1]. Three years later, the present authors developed a simple sheme that evaluated some classical water force fields from the point of view of the extent they can be made consistent with measured diffraction data of liquid water: TIP4P/2005 has proven to be one of the best performing one [2]. Vega and Abascal later performed more extensive tests, involving many thermodynamic and dynamic properties [3]. The wide range of computer simulation studies of Prof. Vega has been extended to other hydrogen bonded systems, like liquid methanol: for this material, he also proposed a potential function, OPLS/2016 [4]. The present study is focused on structural properties and their pressure dependence of computer simulated neat liquid methanol, produced by selected 3-site interatomic potential models, including OPLS/2016.

Methanol (methyl-alcohol, $CH_3$-OH) is the simplest organic molecule that possesses both a hydrophylic (the hydroxil, -OH, group) and a hydrophobic (the methyl, $CH_3$-, group) part. Via the hydroxil groups, methanol molecules are able to form hydrogen bonds (H-bonds), similarly to water. Methanol is interesting from a chemical physics perspective as it can be taken as an analogue to water, in which a small hydrophobic group replaces a single proton. A methanol molecule can form 3 H-bonds (2 acceptors, 1 donor -- and not 4 as water) with its neigbours. Such a difference can cause a significant differences in terms of the topology of the hydrogen bonded network. It is well established that water molecules form a three dimensional network of tetrahedrally coordinated molecules (see, e.g., Refs. [5,6]), whereas in liquid methanol we can detect mainly (branched) chains, and only a very small number of cyclic entities (see, e.g., Ref. [7]). The importance and challenging properties of pure liquid methanol may be emphasised by noting that rather sophisticated computer simulation studies appear continuously, like the very recent, detailed ab initio molecular dynamics investigation of Blach et al. [8].

Water is well known its many anomalies that are generally thought to be a direct consequence of the extended 3D hydrogen bonded network [9, 10]. For instance, the work of Ichiye et al. [11] reports on molecular dynamics simulations that show the anomalous pressure dependence of the self-diffusion coefficient. As methanol closely resembles to water, it is of interest to investigate hydrogen bonded aggregates in simulated methanol as function of pressure. It is established that pure methanol does not form a percolated H-bonded network under ambient conditions, but at very low temperatures, signatures of percolation have been observed [12]. Here we wish to see if pressure can influence the aggregation behaviour of molecules in a similar way in neat methanol.

In the present work, the focus is put on hydrogen bonding properties and their variation with pressure. Pressure dependence is an important aspect for multiple reasons, for instance (1) it is a direct test of the appropriateness of repulsive part of interatomic potentials, (2) recently, and perhaps surprisingly, methanol under extreme conditions has become frequently investigated in the field of astronomy [13].

Concerning the microscopic structure, our primary focus here, diffraction measurements can provide the most direct experimental information. Working under high pressure poses large difficulties in experimental physics in general: this is the main reason why only a few sets of neutron and X-ray diffraction data in (or at least, approaching) the GPa region are available in the literature [14,15]).

Classical computer simulations of pure methanol with interatomic potentials have a rather long history (see, e.g., Ref. [16]). There are two basic types of interactions potentials for this six-atom molecule. (1) The 'all atom' (AA) types consider an interaction site for each atom, so that these are 6-site models whose prototype is the OPLS-AA force field [17]. (2) 'United atom' (UA) types represent methanol molecules by three sites, hydroxyl-H (H, in short), hydroxyl-O (O) and 'C', that unites the methyl-group in one site (see, e.g., Ref. [16]. The use of UA force fields is much cheaper computationally (for an example of their use, see, e.g., Ref [18]), therefore much larger systems may be considered than when all-atom potentials are applied. This was the main reason why it was thought that a comprehensive study on the applicability of UA type potentials would be timely. Their behavior has been particularly poorly explored under pressure, and structural aspects have also not been considered in detail.

A comparison of many potential functions for methanol is reported in Ref. [19]: from the present set, only the TraPPE force field [20] appears there, and no pressure dependence is considered. May we also note that the close agreement between measured X-ray and neutron diffraction data sets shown in Figure 7 of Ref. [19] do not match results published earlier (cf., e.g., Refs. [21,22]. It is worth mentioning, however, that Khasawneh et al. apply the same approximation for the 'scattering power' of the methyl group as was used in [23].

Pressure dependence of H-bond lifetimes has been considered by the simulation study of Wick and Dang [24], who found that increasing pressure increases the duration over H-bond are maintained. This issue is considered here in some detail.

In the present study, we concentrate on three united atom type potentials for methanol, TraPPE [20], UAM-I [25] and the OPLS/2016 [4] parametrisations. The principle is the same for these models: the 4 atoms of the methyl ($CH_3$-) group are 'united' into one 'hydrophobic' interaction site, whereas the hydroxyl O and H remain separate. Interactions contain two components: partial charges and Lennard-Jones parameteres. Hydrogen bonding can occur via the O and hydroxylic H atoms. The philosophy behind their construction, however, is different:
   (i) In the TraPPE force field, the values of the partial charges were borrowed from the OPLS−UA force field [26]. The Lennard−Jones well depth and size parameters for the new interaction sites were determined by fitting to the single-component vapor−liquid-phase equilibria of a few selected model compounds.
   (ii) The main puprose of the introduction of UAM-I (also denoted as 'UAMI-EW' in Ref. [25]) has been to reproduce liquid-vapor and liquid-liquid equilibrium experimental data as close as possible.
   (iii) The three sites of the OPLS/2016 are located at the same positions as those used in the original OPLS model, introduced by Jorgensen [26]. Partial charges and Lennard-Jones parameters were modified by fitting to a selected set of target properties, e.g., solid-fluid experimental thermodynamic data (for details, see Ref. [4]).

The primary goal here is to explore the pressure dependent properties, particularly structural and H-bond related ones: for some reason, these areas have been only scarcely investigated previously for united atom type force fields.

## Computational Methods

Classical molecular dynamics simulations have been conducted in the isothermal-isobaric (NpT) ensemble on liquid methanol, using three 'united atom' type force fields mentioned previously.

The interaction potential between all atoms and/or groups is assumed as a sum of Lennard-Jones (LJ) and Coulomb contributions. Lorentz-Berthelot combination rules are used to determine cross parameters for the relevant potential well depths and diameters.

Molecular dynamics computer simulations of neat liquid methanol have been performed in the isothermal - isobaric (NPT) ensemble at given pressure values, at room temperature. The GROMACS software [27], version 5.1.2, has been used for the all calculations. The simulation box in each run was cubic, the number of molecules in all cases was fixed at 3000. Periodic boundary conditions were applied. Temperature and pressure control has been provided by the V-rescale thermostat and Parrinello-Rahman barostat with $\tau_T$ = 0.5 ps and $\tau_P$ = 2.0 ps. The timestep was 2 fs when calculating static properties, whereas during the shorter runs for determining dynamic properties, it was lowered to 1 fs. Compressibility was set at the value of 4.5e-5 bar$^{-1}$.

The non-bonded interactions were cut-off at 1.4 nm, whereas the long-range electrostatic interactions were handled by the particle mesh Ewald method [28,29] implemented in the GROMACS software package (fourth order, Fourier spacing equal to 0.12), with a precision of $10^{-5}$. The van der Waals correction terms to the energy and pressure were applied. In order to maintain the geometry of methanol intra-molecular bonds rigid, the LINCS [30] algorithm was used.

Calculation of the total scattering structure factors (TSSF), is a non-trivial excersise when ... after various attempts, we ended up with the simplest approximation in the neutron case:

$$b_{(CD3)} = b_{(C)} + 3*b_{(D)}, \qquad (1)$$

where 'b' are the scattering lengths for neutrons ($b_{(C)}$=6.65 fm, $b_{(D)}$=6.67 fm, $b_{(O)}$=5.80 fm). This approximation has already been employed during a pilot molecular dynamics simulation of propanol-water liquid mixtures [23].

The total structure scattering structure factor, *F(Q)*, that can be obtained by X-ray and neutron diffraction experiments, may be calculated from the partial rdf's according to the equation

$$F(Q) = \sum_{i \geq j} \frac{(2 - \delta_{ij}) x_i x_j f_i(Q) f_j(Q) S_{ij}(Q)}{(\sum_{i=1}^{n} x_i f_i(Q))^2} \qquad (2)$$

Here, $f_i$ is the scattering length (for neutron diffraction) or scattering form factor (X-ray diffraction) of atom type *i* (that depends on *Q* in the case of X-ray diffraction, and is constant in the case of neutron diffraction), *n* is the number of scattering site (atom/united atom) types, and $x_i$ is the corresponding mole fraction. $S_{ij}(Q)$, the partial structure factor (PSF), is defined from the partial radial distribution functions (PRDF), $g_{ij}(r)$, according to the following equation:

$$S_{ij}(Q) = 4\pi \rho_j \int_0^{r_{max}} r^2 \left( g_{ij}(r) - 1 \right) \frac{\sin(Qr)}{Qr} dr \qquad (3)$$

where $\rho_j$ is the atomic(/site, cf. united methyl group) number density of species '*j*'.

For characterising hydrogen bonding properties, one needs to define hydrogen bonds between neighbouring methanol molecules. Throughout this work we use one of the popular definitions for the hydrogen bond (see, e.g. in Ref. [31], also applied in Refs. [12, 22]). According to the purely geometric definition, two molecules are considered to be hydrogen bonded to each other if they are found at a distance r(O···H) < 2.5 Å, and the O...OH angle is smaller than 30 degrees. When using the energetic definition, two molecules were considered to be hydrogen bonded if (1) they were found at a distance r(O···H) < 2.5 Å and (2) the interaction energy was less than - 3.0 kcal/mol. It has been found that the exact H-bond definition applied has not resulted in significant differences concerning the main conclusions, particularly when trends are more important than individual values. Throughout the rest of this study, results for the energetic definition are shown and discussed.

Collective structural properties have been calculated by an in-house software that is based on the work of Chihaia et al. [32]. Out of the various characteristics that can be determined from simulation trajectories (see Ref. [33]), here the potential model, as well as the pressure, dependence of cluster size distributions and ring size distributions are displayed.

Two molecules are regarded as belonging to the same cluster if a connection, via a chain of hydrogen bonds, can be found between them. A system is said to be percolating if the number of molecules in the largest cluster is in the order of the system size. According to previous studies (e.g., Refs. [7,12]), percolation does not occur in pure alcohols under ambient condidions. However, cluster size distributions, that have been calculated as described in, e.g., [12, 33], may prove to be a powerful tool for detecting pressure effects on collective structural features.

Rings are a particular kind of H-bonded clusters: the criterion is that there must be a path in the cluster which returns to the molecule assigned as origin. Here, primitive rings have been identified. as described in, e.g., Refs. [33,34]: a ring is called 'primitive' if it cannot be decomposed into smaller rings.

Turning now to dynamic properties, self-diffusion coefficients and reorientational characteristic times have been calculated as described in Ref. [35].

The self-diffusion coefficient (*D*) was estimated using the Einstein- Smoluchowski relation, from the mean squared displacements (MSD) of the centres of mass of water and ethanol molecules:

$$D = \lim_{t \to \infty} \frac{1}{6Nt} \sum_{i=1}^{N} <r_i(t) - r_i(0)>^2 \qquad (4)$$

where $r_i(t)$ and $r_i(0)$ are the positions of the centres of mass of methanol molecules at time *t* and *0*, respectively, and the <...> denotes an ensemble average. The effect of using every *n*-th saved configuration (*n* = 1, 5, 20) during the MSD calculation was negligible, as it has been shown by Gereben et al. [36] for simulated liquid water (we note that in Ref. [36], a relatively detailed discussion is available on the calculation of the self-diffusion coefficient).

Reorientational dynamics can be characterized via autocorrelation functions of unit vectors connected to well-defined molecular axes, such as along the OH-bond, or the perpendicular one to the COH methanol molecular plane. The characteristic decay times, that is the integral of the autocorrelation function of the unit vectors mentioned above may be measurable using NMR experiments [37].

Hydrogen bond lifetimes have been calculated via the simplest concept: we determine the distribution of the lifetimes of individual H-bonds and report their average (for a more detailed discussion, see, e.g., Ref. [38]). A critical distinction in MD lifetime calculations is between "continuous" and "intermittent" lifetimes (see, e.g., Ref. [39]). The continuous lifetime requires the hydrogen bond to be present at every single frame from $t_0$ to $t_0+\tau$, where $t_0$ is the time of origin for a particular H-bond and is the time during which the same H-bond exists. In contrast, the intermittent lifetime allows for small, transient breaks in the hydrogen bond (up to a specified number of frames) while still considering it "present" over the longer duration. This intermittency parameter can significantly alter the calculated lifetime, with higher intermittency values generally leading to longer reported lifetimes. Here we report both continuous and intermittent lifetimes (see below), as a function of the force field, and of pressure.

## Results and Discussion

**Densities resulting from isothermic-isobaric molecular dynamics simulations**
Arguably the most basic test of a potential function is whether it can reproduce the experimentally determined densities. For this reason, this quantity has been calculated for the UAM-I, TraPPE and OPLS/2016 force fields, as a function of pressure. Figure 1 compares simulation results with available experimental data [40] up to 4 kbar. Comparison to experimental data is rather favourable for each force field considered here. This finding provides a firm basis for more detailed investigations.

Scrutinizing the behaviour of the curves, a small but systematic difference between the applied force fields becomes apparent: density values originating from the UAM-I potential follow experimental data above ca. 1.5 kbar more closely than those from the other two potentials. On the other hand, below ca. 1 kbar, the OPLS/2016 force field provides the best density estimates. Although not too important from the practical point of view, this small detail exemplifies why pressure-dependent investigations are important: ab ovo, there is no guarantee that a force field that is appropriate over a given pressure domain would be as good over other pressure domains. In the realm of computer simulations, the only way to find this out is 'trial and error'. The present study, although this not the main focus here, provides information on which potential function is more suitable, with respect to density, in certain pressure domains.

Following this line a little further, it should be noted that the sophisticated all-atom potential function applied in the MD simulations of Wick and Dang [24] yields a significantly worse agreement with experimental data [40] than any of the united atom force fields considered here. That is, since density is the most basic poroperty of a system, predictions made in Ref. [24] should be taken with some caution.

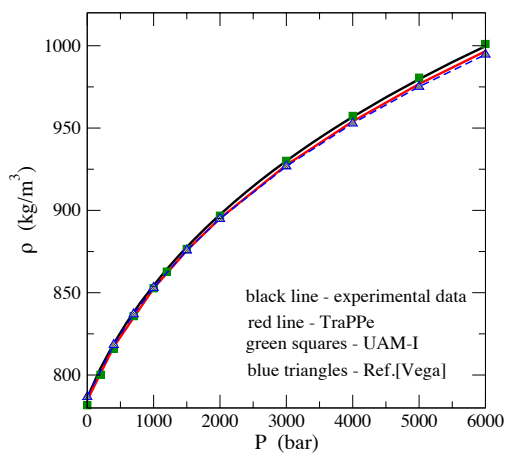

Figure 1. Comparison of simulated and experimental density values for the three united atom type potential models considered here, up to 6 kbar. Experimental data are from Ref. [40] (black line). Red line: TraPPE, green squares: UAM-I, blue triangles: OPLS/2016.

**Two particle correlations: total scattering structure factors and partial radial distribution functions**

In Figure 2, experimental (from Ref. [15]) and simulated (present work) neutron weighted total scattering structure factors are shown. All the three simulated TSSF-s follow the major experimental trend with increasing pressure, i.e., the well observable shift of the position of the first maximum towards higher $Q$ values, as well as the significant increase of the height of the first maximum. Although experimental curves could not be reproduced in detail, it is worth noting that the level of agreement with experiment is comparable to what was found in Ref. [15] for an all-atom type interaction potential. This is a strong indication for that united atom force field are competitive from the point of view of predicting major structural trends in liquid methanol on increasing pressure.

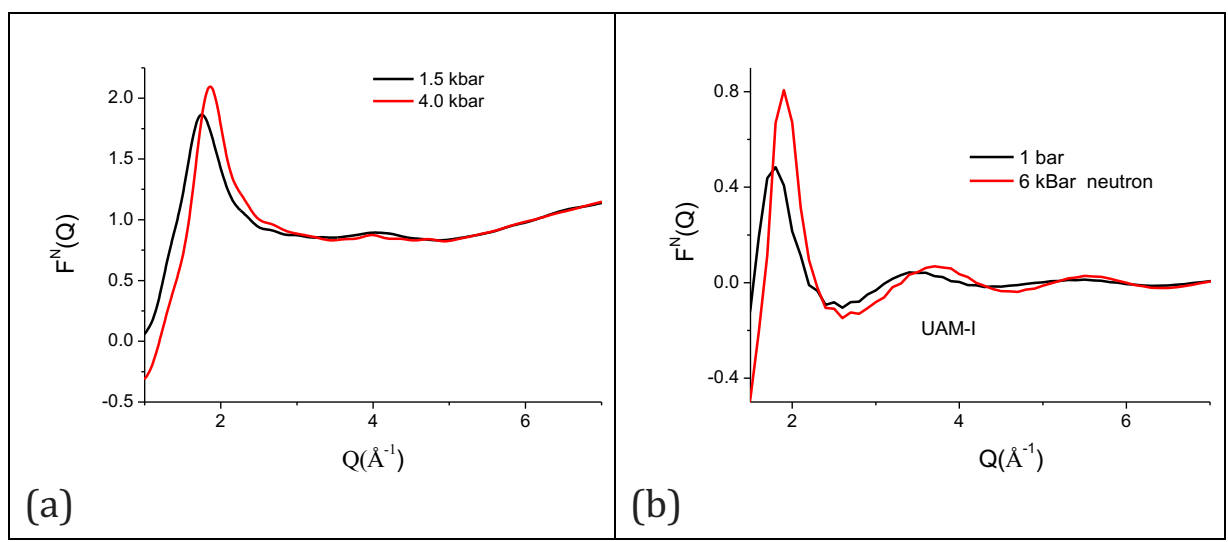

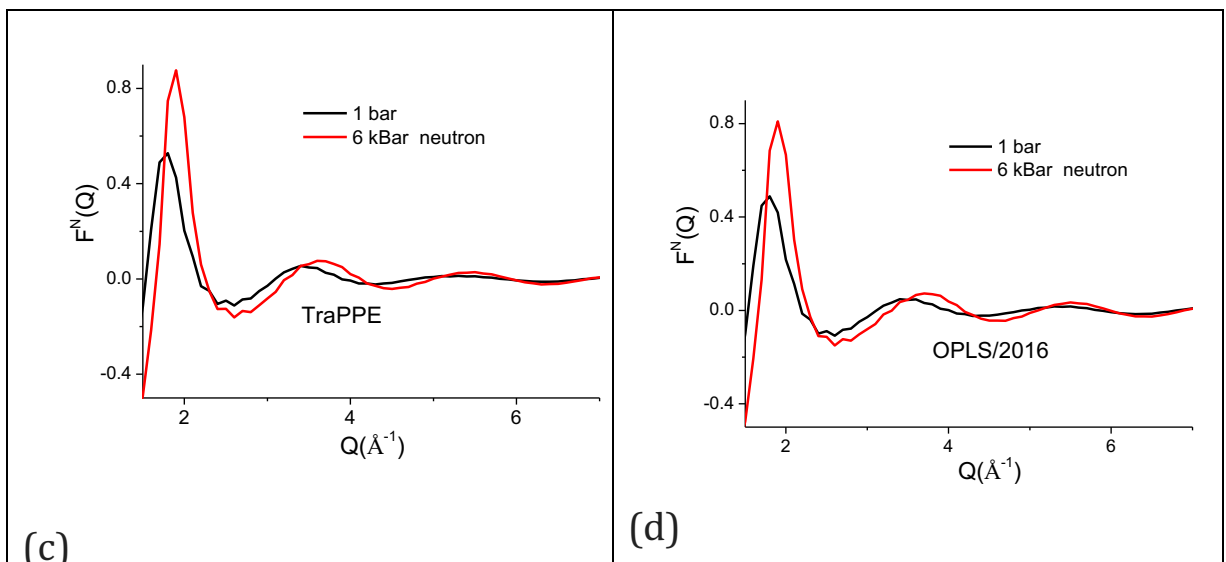

(c) (d)

Figure 2. (a) Experimental neutron diffraction data of liquid methanol at 1.5 kbar (black) and 4 kbar (red). (Note that during the experiments, the lowest pressure in the aluminium pressure cell, due to technical reasons was 1.5 kbar.) (b) Intermolecular total scattering structure factors, as calculated from MD simulations for the UAM-I potential for 1 bar (black) and 6kbar (red). (c) Same as (b), but showing simulation results for the TraPPeEpotential. (d) Same as (b), but showing simulation results for the OPLS/2016 potential.

Figure 3 displays the X-ray (part (a)) and neutron (part (b)) weighted total intermolecular total structure factors at the highest pressure (6 kbar) considered here. (Note that the the neutron weighted total was calculated by using weights calculated for the deuterated sample, $CD_3OD$, cf. Ref. [15].) Interestingly, in the X-ray case the OPLS/2016 potential brings about a well distinguishable curve, whereas the neutron weighted TSSFs are almost identical. This difference is caused by the different weghting factors of the partial functions: the O-O partials have a much bigger contribution in the X-ray case, whereas calculated neutron data are dominated by the huge 'united atom' scattering length of the (deuterated) methyl group ($CD_3$-).

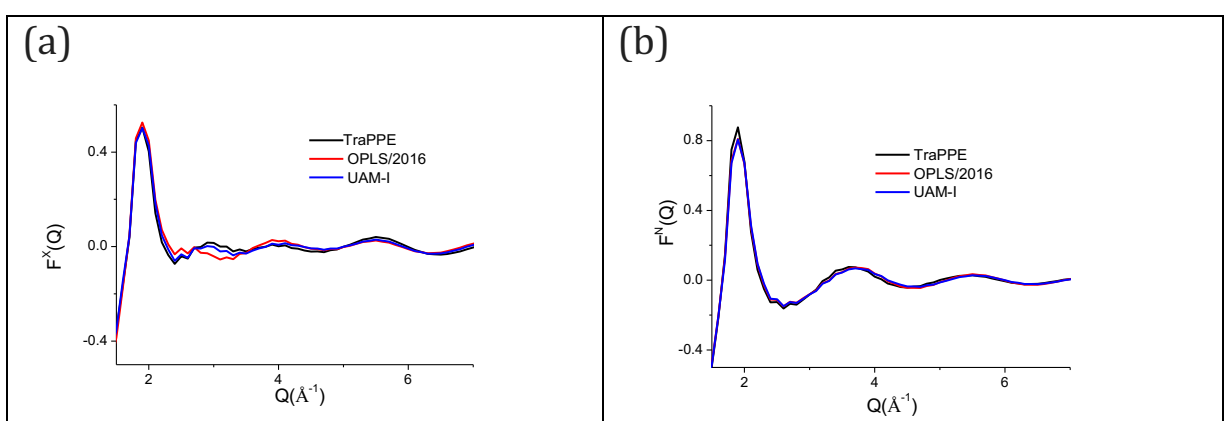

Figure 3. (a) X-ray and (b) neutron weighted total scattering structure factors for the three potentials at 6 kbar. Note the easily distinguishable X-ray weighted function corresponding to the OPLS/2016 force field.

Figure 4 compares partial radial distribution functions resulting from the different force fields at 1 bar (part (a)) and at 6 kbar (part (b)). The largest alterations can be detected for the OO and OH PRDFs, at both pressure values. This is an indication of that the developers of the three united atom force fields have varied the 'hydrophylic' parameters over a broader range, whereas concerning the 'hydrophobic' parts, there seems to be a concensus. Concerning the CC PRDF, it is the TraPPE potential that produces results slightly different from the other two, whereas for the CO PRDF, the OPLS/2016 potential can be distinguished the most clearly.

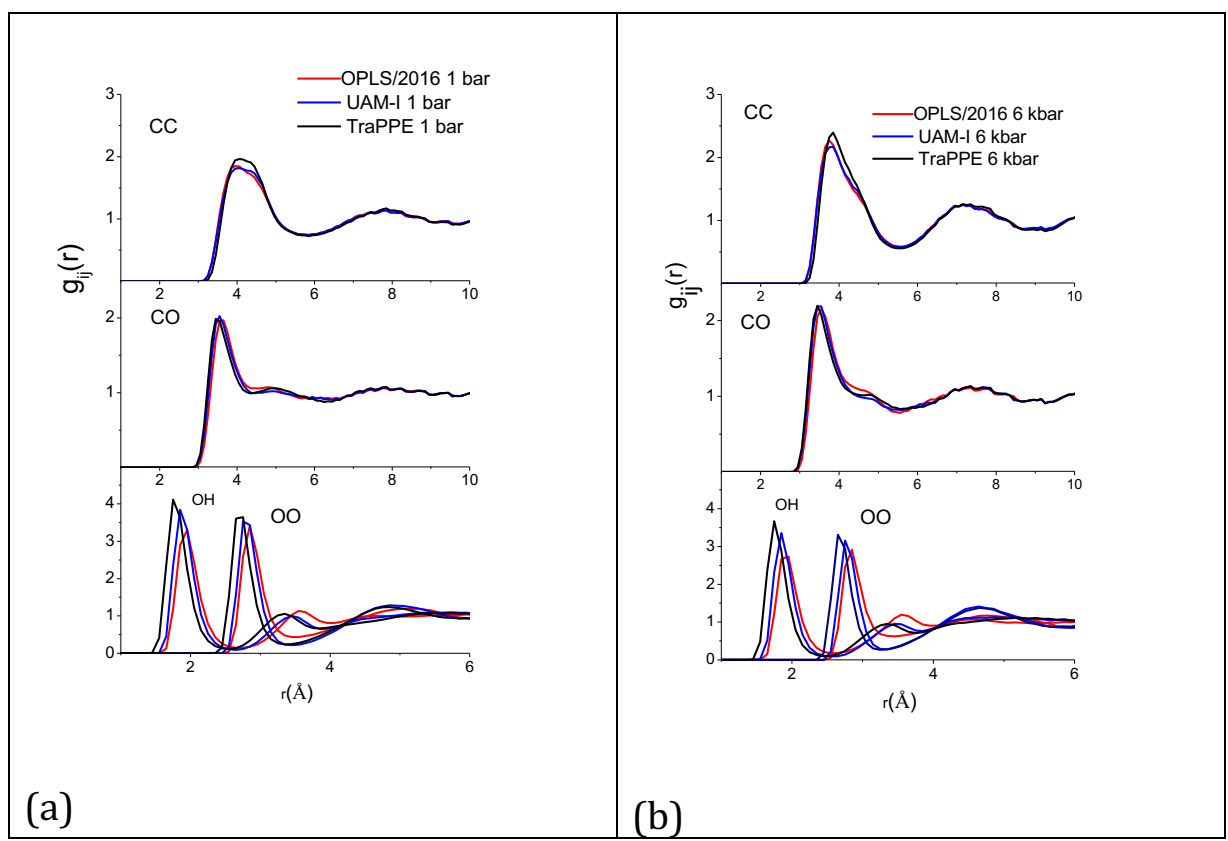

Figure 4. Partial radial distribution functions at (a) 1 bar and (b) 6 kbar for the three UA-type potentials. Note the behaviour of the OO (and OH) functions originating from the OPLS/2016 force field.

In Figure 5, some of the partial structure factors (PSF) from the TraPPE, UAM-I and OPLS/2016 force fields are compared at pressure values of 1bar and 6 kbar. As it could be expected from the comparison of the PRDF-s (cf. Figure 4), the largest differences between the force fields show up on the OO PSF, the curves from the OPLS/2016 potential being the most distinct. This explains the deviations seen on the X-ray weighted total structure factors (Figure 3).

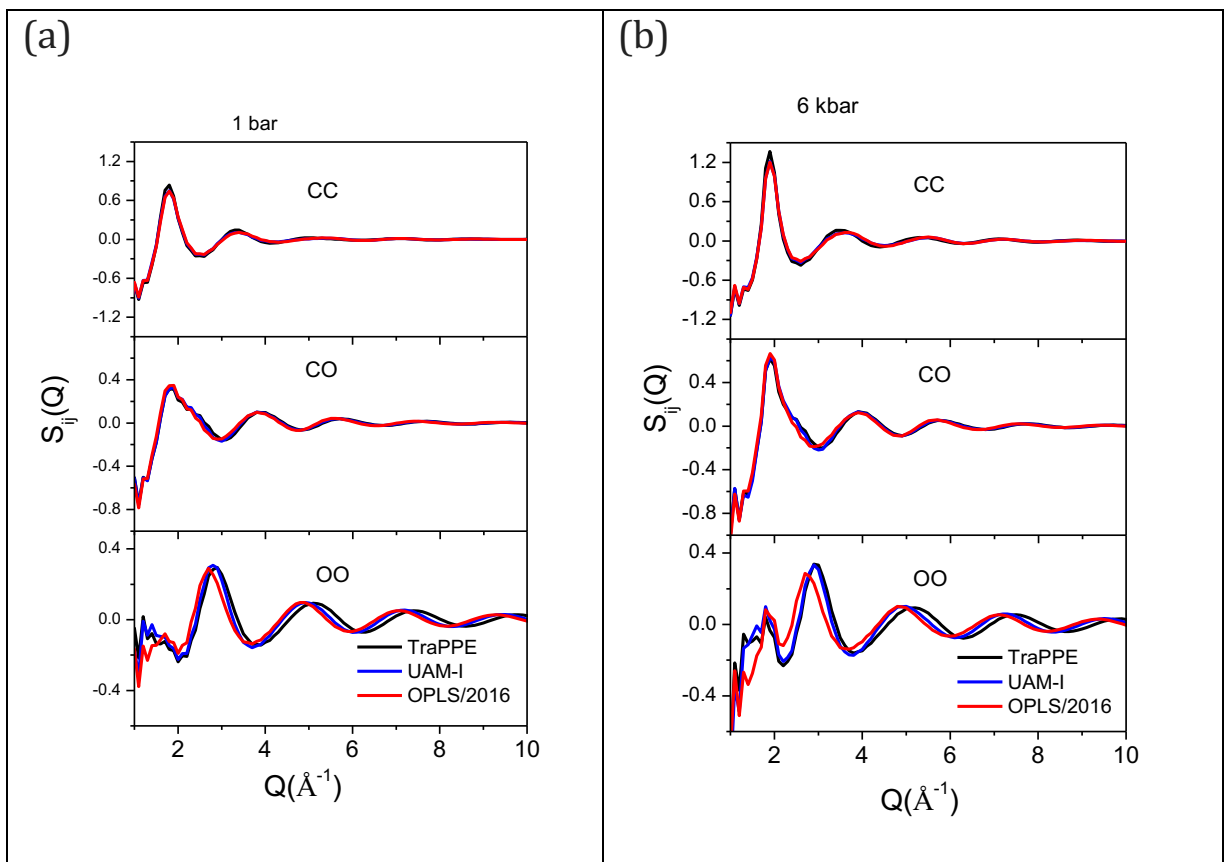

Figure 5. Comparison of partial structure factors predicted by the three force fields comsidered here at (a) 1 bar, and (b) at 6 kbar. (The OH PSF-s are not shown, since their contribution to the TSSFs are nearly negligible.)

Figure 6 compares partial structure factors at 1 bar and at 6 kbar for each of the three UA-type potentials considered here separately: this way, the effects of increasing pressure may be spotted easily. Note that the shift of the maxima, present on the TSSFs (cf. Figure 3) is only apparent on the CC and OO PSFs, whereas the CO PSF remains nearly intact in this respect. On the other hand, the increasing intensity of the first maximum, another observable distinction on the TSSFs, can be detected on the CC and CO PSFs. These observations can explain the deviations seen on the X-ray weighted total structure factors (Figure 3).

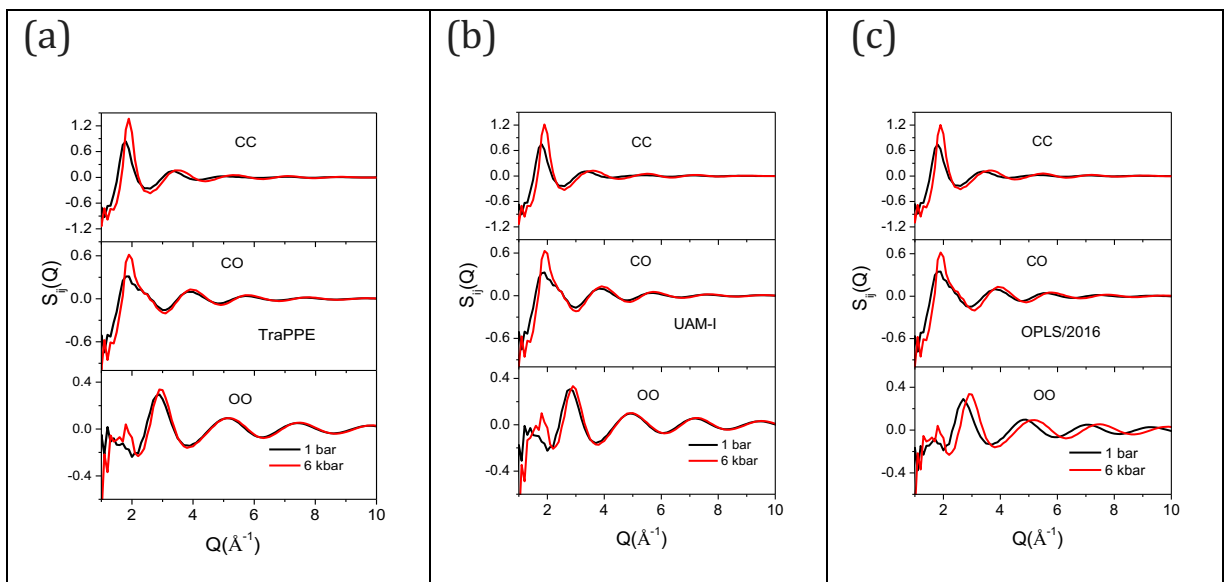

Figure 6. Comparison of partial structure factors at 1 bar (black lines) and 6 kbar (red lines) coming from simulations using the (a) TraPPE, (b) UAM-I and (c) OPLS/2016 force fields.

**Collective properties involving more than two molecules: the influence of pressure**

Perhaps the simplest way to characterize hydrogen bonded liquids is calculating the distribution of the number of hydrogen bonded neighbours. Table 1 provides numerical data, so that the minute differences observed can be spotted more easily.

Inspecting Table 1, it is apparent that neither the potential model chosen, nor the increasing pressure has a serious influence on distribution of the number of hydrogen bonded neighbours. Far the most abundant are the two-fold coordinated molecules and there are only very few methanol molecules that are 'monomers', without a single H-bonded neighbours. From the minute differences it appears from the MD simulations using the OPLS/2016 force field at a pressure of 6 kbar produce the highest proportion of 2-fold coordinated molecules. OPLS/2016 is the only force field where the proportion of twofold H-bonded molecules grow with pressure, thus making the distribution the sharpest of the three UA-type potential models considered here. It is worth noting that data in Table 1 are consistent of pressure dependent Raman scattering results of Arencibia et al. [41].

Table 1. Distribution of the number of hydrogen bonded neighbours in liquid methanol, at pressure values of 1 bar and 6 kbar. Values are given in fractions.

| Number of H-bonds | OPLS/2016 | | UAM-I | | TraPPE | |
|---|---|---|---|---|---|---|
| | 1 bar | 6 kbar | 1 bar | 6kbar | 1 bar | 6 kbar |
| 0 | 0.012 | 0.008 | 0.011 | 0.009 | 0.012 | 0.009 |
| 1 | 0.160 | 0.147 | 0.159 | 0.161 | 0.165 | 0.164 |
| 2 | 0.803 | 0.809 | 0.777 | 0.768 | 0.766 | 0.761 |
| 3 | 0.025 | 0.035 | 0.054 | 0.062 | 0.056 | 0.065 |

Another simple function that correlates well with nearest neighbors according to the O..H intermolecular distance is the distribution of O…OH angles (the H-bonding angles). Argueably, these are the simplest many body correlation functions. Force field dependent O…OH angle distribution functions are shown in Figure 7 for liquid methanol at 1 bar and 6 kbar.

It can be concluded that, surprisingly, neither the positions, not the height of the maxima change significantly with pressure. On the other hand, the interatomic potential applied does have a significant effect: at both 1 bar and 6 kbar, the OPLS/2016 force field produces significantly different distributions, both in terms of the position and the height of the maximum. As it can be devised from Figure 7, this potential model allows for H-bond angles that deviate, within the limits set by the definition of H-bonds (see above), more from the regular straight angle than the hydrogen bond angles brought about by the TraPPE and UAM-I models.

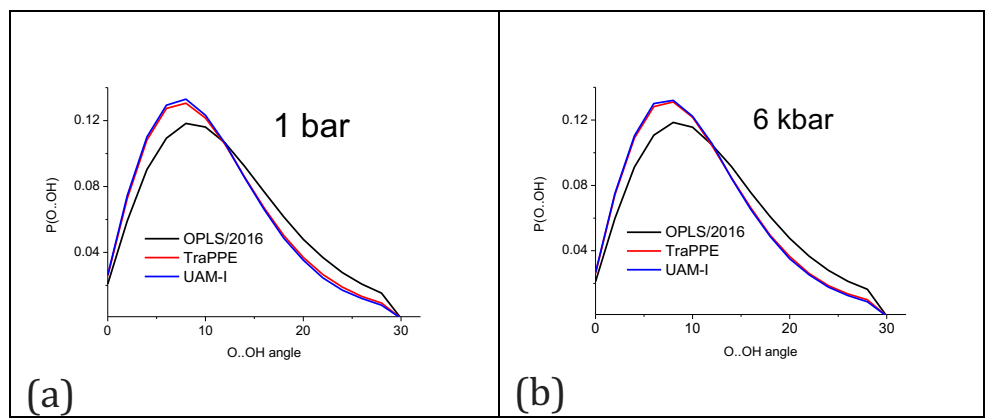

Figure 7. Distribution of the hydrogen bonding angles in liquid methanol, at (a) 1 bar and (b) 6 kbar. Note that the distribution belonging to the OPLS/2016 force field is significantly different at both pressure values.

As it is apparent by looking at Figure 8, the distribution of the size of cyclic entities is perhaps the property that is the most sensitive to the actual force field, as well as to pressure. Although the concentration of rings is rather small, between 10 to 20 in a configuration of 3000 molecules, clearly OPLS/2016 at 6 kbar produces the smallest, whereas TraPPE at 1 bar the largest number of cyclic aggregates. Increasing pressure consistently brings about a decrase in terms of the number of rings. The most abundant rings contain 5 molecules in each case – note, however, that in the OPLS/2016 simulation at 1 bar, 6-membered cycles are nearly as frequent. These values are consistent with findings of Reverse Monte Carlo [42] calculations of Vrhovsek et al. [7], as well as the very recent ab inition molecular dynamics simulations of Blach et al. [8].

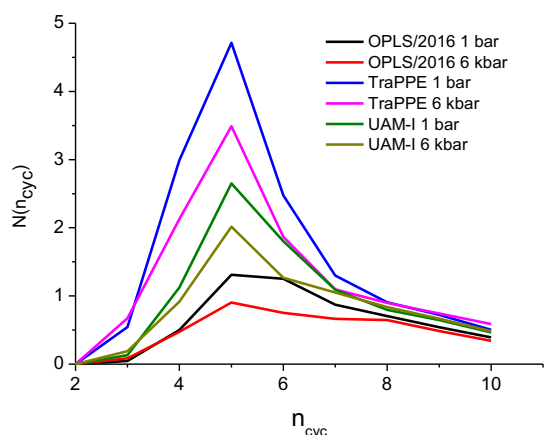

Figure 8. Ring size distributions for the three force fields considered, at 1 bar and 6 kbar. '*x*' axis: the number molecules in a ring; '*y*' axis: the average number of cyclic entities of size $n_{cyc}$ in a particle configuration (containing 3000 methanol molecules). The concentration of rings is, indeed, very low.

Figure 9 reports on cluster size distributions: here only the number of H-bonded methanol molecules in an aggregate matter, irrespective of the shape of the aggregates. Although the effects are not huge, increasing pressue causes an increase of the sizes of clusters held together by hydrogen bonds. This effect is smallest for the UAM-I force field, whereas it is largest for the OPLS/2016 one. The OPLS/2016 force field consistently produces the smallest aggregates, as well as the lowest number of aggregates of a given size.

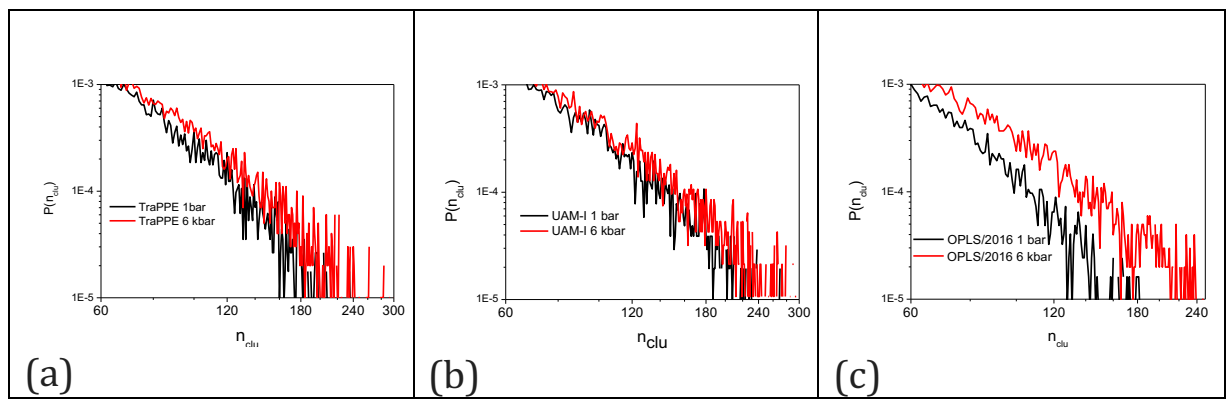

Figure 9. Pressure dependence of cluster size distributions in liquid methanol, (a) TraPPE, (b) UAM-I, (c) OPLS/2016.

As a brief summary on collective structural features, it is fair to state that pressure profoundly influences both the number of cyclic structures and the size of H-bonded clusters: the latter grow on the expense of the former.

**Dynamic properties**
*Self diffusion coefficients* of the molecular centres have been calculated as described in Refs. [35,36]. Numberical data are provided in Table 2. Experimental values, as determined from NMR results [43], are also provided where available.

Table 2. Pressure dependence of the self diffusion constant in liquid methanol, in units of $10^{-8}$ m$^2$/s. Experiental data are from Ludwig et al. [43].

|  | OPLS/2016 | UAM-I | TraPPE | exp |
|---|---|---|---|---|
| 1bar | 0.327 | 0.251 | 0.254 | 0.230 |
| 2kbar | 0.157 | 0.128 | 0.119 | 0.140 |
| 4 kbar | 0.093 | 0.077 | 0.077 | |
| 6 kbar | 0.060 | 0.046 | 0.047 | |

Increasing pressure decreases *D*, as it has already been observed in Ref. [44]. Overall, OPLS/2016 predicts higher values than the other two force fields. At higher pressure, OPLS/2016 is at least as accurate as UAM-I and TraPPe, while at 1 bar, UAM-I and TraPPE provides more accurate values.

We note that the work of Wick and Dang [24] also predict decreasing *D* values with increasing presssure. However, their numerical values differ from experimental data significantly (as it is admitted by those authors themselves).

*Reorientational times* have been calculated as described in Ref. [35]. Numerical data of the various characteristic times are given in Table 3.

Table 3. OH1, OH2 and COH plane relaxation times in ps. Experimental data are from Ref. [45].

|  | OPLS/2016 | | | UAM-I | | | TraPPE | | | |
|---|---|---|---|---|---|---|---|---|---|---|
|  | $t_1$ | $t_2$ | $t_{perp}$ | $t_1$ | $t_2$ | $t_{perp}$ | $t_1$ | $t_2$ | $t_{perp}$ | exp. |
| 1bar | 12.05 | 5.29 | 5.03 | 15.92 | 6.76 | 7.41 | 15.15 | 6.58 | 7.87 | 5.00 |
| 2 kbar | 16.39 | 7.75 | 6.58 | 21.74 | 9.71 | 10.56 | 20.41 | 9.62 | 10.75 | |
| 4 kbar | 20.00 | 10.00 | 9.35 | 27.03 | 12.99 | 13.16 | 25.64 | 12.66 | 13.33 | |
| 6 kbar | 25.00 | 12.66 | 11.90 | 32.26 | 16.13 | 16.95 | 31.25 | 16.39 | 16.95 | |

All relaxation times grow with pressure monotonously. The very limited experimental data is reproduced best by the OPLS/2016 force field (it is not intended, however, that any assessment on the force fileds would be based on this limited comparison).

*Lifetimes of hydrogen bonds* have been calculated with ('t1' in the tables below) and without ('t1*' in the tables below) of an intermittence time of 0.15 ps (for details of the method of calculations, please see above and Ref. [38]). Results are summarized in Table 4.

Table 4. Hydrogen bond lifetimes from each potential models at 1 bar and 6 kbar (in ps).

| TraPPE | t1 | t1* | UAM-I | t1 | t1* | OPLS/2016 | $t_1$ | $t_1$* |
|---|---|---|---|---|---|---|---|---|
| 1 bar | 5.17 | 0.43 | 1 bar | 5.92 | 0.46 | 1 bar | 3.46 | 0.44 |
| 2 kbar | 4.75 | 0.39 | 2 kbar | 5.41 | 0.42 | 2 kbar | 2.97 | 0.40 |

| 4 kbar | 4.47 | 0.37 | 4 kbar | 5.05 | 0.39 | 4 kbar | 2.81 | 0.37 |
| 6 kbar | 4.26 | 0.35 | 6 kbar | 4.59 | 0.37 | 6 kbar | 2.51 | 0.37 |

Note the difference between intermittent times 0.15 and 0.0 ps. For the former, OPLS/2016 is markedly different. For the latter, the results for the three force fields are basically identical. This suggests that librational motions induced by the OPLS/2016 force field are different from those resulting from UAM-I and TraPPE. For each potential model, and for both the 'continuous' and the 'intermittent' concepts, H-bond lifetimes decrease with increasing pressure.

The above finding is in contradiction with simulation results presented in Ref. [24]. Note, however, that both their [24] estimates of density and the diffusion coefficient fail to reproduce experimental data as closely as the present work. Both quantitities are related to hydrogen bonding. Moreover, trends reported by Wick and Dang [24] for both properties are noticeably different from the experimental ones: i.e., the present predictions of H-bond lifetimes may be considered more reliable those published in Ref. [24]. However, quite clearly, further work is necessary to clarify the issue of the pressure dependence of hydrogen bond lifetimes.

**Conclusions**

In summary, new classical molecular dynamics simulations of neat liquid methanol have been conducted, at pressure values of 1 bar and 2, 4 and 6 kbar, using three united atom type force fields. These values provide overlap with quite a few experimentally measurable quantities. Concerning the pressure dependent structure and dynamics, the following conclusions may be drawn:

(i) A general observation is that while individual/pair properties are not influenced by pressure too strongly, pressure effects on collective properties, like ring- and cluster size distributions, are significant.

(ii) UAM-I and TraPPE provide similar values in terms of almost all properties reported, whereas results from OPLS/2016 most frequently behave differently. The largest spread between the three force fields has been found for the distribution of cyclic entities.

(iii) Each potential provides qualitative agreement with measured total scattering structure factors. Based on the strong similarity between the neutron weighted structure factors (see Figures 2 and 3), no strong statement can be made concerning which of the force fields stuied here is the 'best'. (Note that X-ray weighted total scattering structure factors from OPLS/2016 behave slightly differently here, again, but this is mainly because the weighting factors of the partial structure factors differ quite much for the two kinds of scattering experiments.)

(iv) Concerning hydrogen bonded aggregates, cluster sizes increase, while the abundance of cyclic entities decreases with pressure. OPLS/2016 exhibits that largest pressure effect, and provides the smallest clusters consistently.

(v) Hydrogen bonding lifetimes decrease with pressure for each of the three UA type potentials investigated here, all of which predict decreasing lifetimes with growing pressure. The intermittent lifetimes vary between ca. 6 and 2 ps, whereas the continuous lifetimes remain in a much narrower region, between 0.46 and 0.35 ps.

A straightforward extension of the present work would be to examine neat higher alcohols (ethanol, 1- and 2-propanol, etc…), as well their mixtures with water. A similar, preliminary account has already appeared for propanol-water mixtures [23].

Another handy development would be to consider more kinds of experimental data to compare with. Dielectric properties, particularly the static dielectric permittivity, offers a logical choice: embryonic attempts have already been made in this direction [46], and the very recent report of Song and Wu [47] on liquid water outlines exiting opportunities.

**Ackowledgements**

This work was supported by the National Research, Development and Innovation Office of Hungary (NKFIH, Grant No. K-142429). Technical support from Magdalena Aguilar at the Institute of Chemistry of the UNAM is gratefully acknowledged. LP is grateful for the opportunity to conduct synchrotron radiation experiments at the BL04B1 high pressure beamline of SPring-8 (Hyogo, Japan), with the approval of the Japan Synchrotron Radiation Research Institute (JASRI)  (Proposal Nos. 2018A1120 and 2018B1382). Without these experiments, the inherent difficulties with high pressure synchrotron X-ray diffraction could not have been fully appreciated.